\documentclass[pdflatex,sn-chicago,iicol]{sn-jnl}

\usepackage{graphicx}%
\usepackage{multirow}%
\usepackage{amsmath,amssymb,amsfonts}%
\usepackage{amsthm}%
\usepackage{mathrsfs}%
\usepackage[title]{appendix}%
\usepackage{xcolor}%
\usepackage{textcomp}%
\usepackage{manyfoot}%
\usepackage{booktabs}%
\usepackage{algorithm}%
\usepackage{algorithmicx}%
\usepackage{algpseudocode}%
\usepackage{listings}%

\theoremstyle{thmstyleone}%
\theoremstyle{thmstyletwo}%

\theoremstyle{thmstylethree}%

\begin{document}

\title[Quantitative Fairness - A Framework For The Design Of Equitable Cybernetic Societies]{Quantitative Fairness - A Framework For The Design Of Equitable Cybernetic Societies}


\author*[1]{\fnm{Kevin} \sur{Riehl}}\email{kriehl@ethz.ch}

\author[1]{\fnm{Michail} \sur{Makridis}}\email{mmakridis@ethz.ch}

\author[1]{\fnm{Anastasios} \sur{Kouvelas}}\email{kouvelas@ethz.ch}

\affil*[1]{\orgdiv{Traffic Engineering Group, Institute for Transport Planning and Systems}, \orgname{ ETH Zurich}, \orgaddress{\street{Stefano-Franscini-Platz 5}, \city{Zürich}, \postcode{8093}, \state{Zürich}, \country{Switzerland}}}


\abstract{
Advancements in computer science, artificial intelligence, and control systems of the recent have catalyzed the emergence of cybernetic societies, where algorithms play a significant role in decision-making processes affecting the daily life of humans in almost every aspect. 
Algorithmic decision-making expands into almost every industry, government processes critical infrastructure, and shapes the life-reality of people and the very fabric of social interactions and communication. 
Besides the great potentials to improve efficiency and reduce corruption, missspecified cybernetic systems harbor the threat to create societal inequities, systematic discrimination, and dystopic, totalitarian societies.
Fairness is a crucial component in the design of cybernetic systems, to promote cooperation between selfish individuals, to achieve better outcomes at the system level, to confront public resistance, to gain trust and acceptance for rules and institutions, to perforate self-reinforcing cycles of poverty through social mobility, to incentivize motivation, contribution and satisfaction of people through inclusion, to increase social-cohesion in groups, and ultimately to improve life quality.
Quantitative descriptions of fairness are crucial to reflect equity into algorithms, but only few works in the fairness literature offer such measures; the existing quantitative measures in the literature are either too application-specific, suffer from undesirable characteristics, or are not ideology-agnostic.
Therefore, this work proposes a quantitative, transactional, distributive fairness framework, which enables systematic design of socially-feasible decision-making systems. 
Moreover, it emphasizes the importance of fairness and transparency when designing algorithms for equitable, cybernetic societies.
}

\keywords{Resource Allocation, Equitable Societies, Distributive Fairness, Procedural Fairness, Algorithmic Fairness}



\maketitle

\section{Introduction}\label{intro}

Technology is increasingly employed for the automation of processes that were previously performed by humans. 
This shift has expanded from early applications in agriculture, manufacturing, and mechanical processes to now encompass planning, decision-making, and control processes~\citep{xu2018fourth}.

Especially the advancements in computer science, artificial intelligence, and control systems of the recent have catalyzed the emergence of cybernetic societies, where algorithms play a significant role in decision-making processes that affect the daily life of humans in almost every aspect. 
This algorithmic decision-making is becoming more prevalent across industries, from finance and healthcare to media, retail and customer service, in the life-reality of citizens of smart and mega cities, and it also involves the design and operations of energy and transportation networks. 
Algorithms even influence the very fabric of our social interactions, personal relationships, and communication using digital media and social networks.
What's more, automated processes are increasingly employed even in law-enforcement, budget allocation, and planning at the governmental level~\citep{friedman2019power,larsson2022technocracy,ashby1956introduction}. 

Automated processes offer numerous potential benefits, such as increased efficiency and objectivity of decisions, improved enforcement of legislation, reduced corruption, acceleration of bureaucratic processes, standardization of processes, and the automation of repetitive tasks that negatively affect mental health of human workers~\citep{abbott2024technocracy}.
At the same time, missspecified or purposefully misused technologies harbor the threat to create societal inequities through systematic discrimination, and enable even more corrupt systems through mass surveillance technology and extreme restriction of individual freedom~\citep{hossain2020explainable}.

Together with efficiency, fairness plays a crucial role when designing and implementing philanthropic, cybernetic systems, that serve and benefit humans~\citep{friedman2019power}.
\begin{itemize}
    \item Only systems that promote equitable societies can guarantee that these cybernetic systems serve people, and not people serving these systems.
    \item The implementation of cybernetic systems in practice often fails due to public resistance and equity concerns~\citep{gu2018congestion}.
    \item Cybernetic systems often support self-coordination of selfish, rational individuals, and to align egoistically optimal with societally optimal outcomes; doing so, fairness is the foundation to achieve any form of cooperation of individuals in large populations~\citep{gurney2021equity}.
\end{itemize}

\begin{figure*}[ht!]
    \centering
    \includegraphics[width=0.9\textwidth]{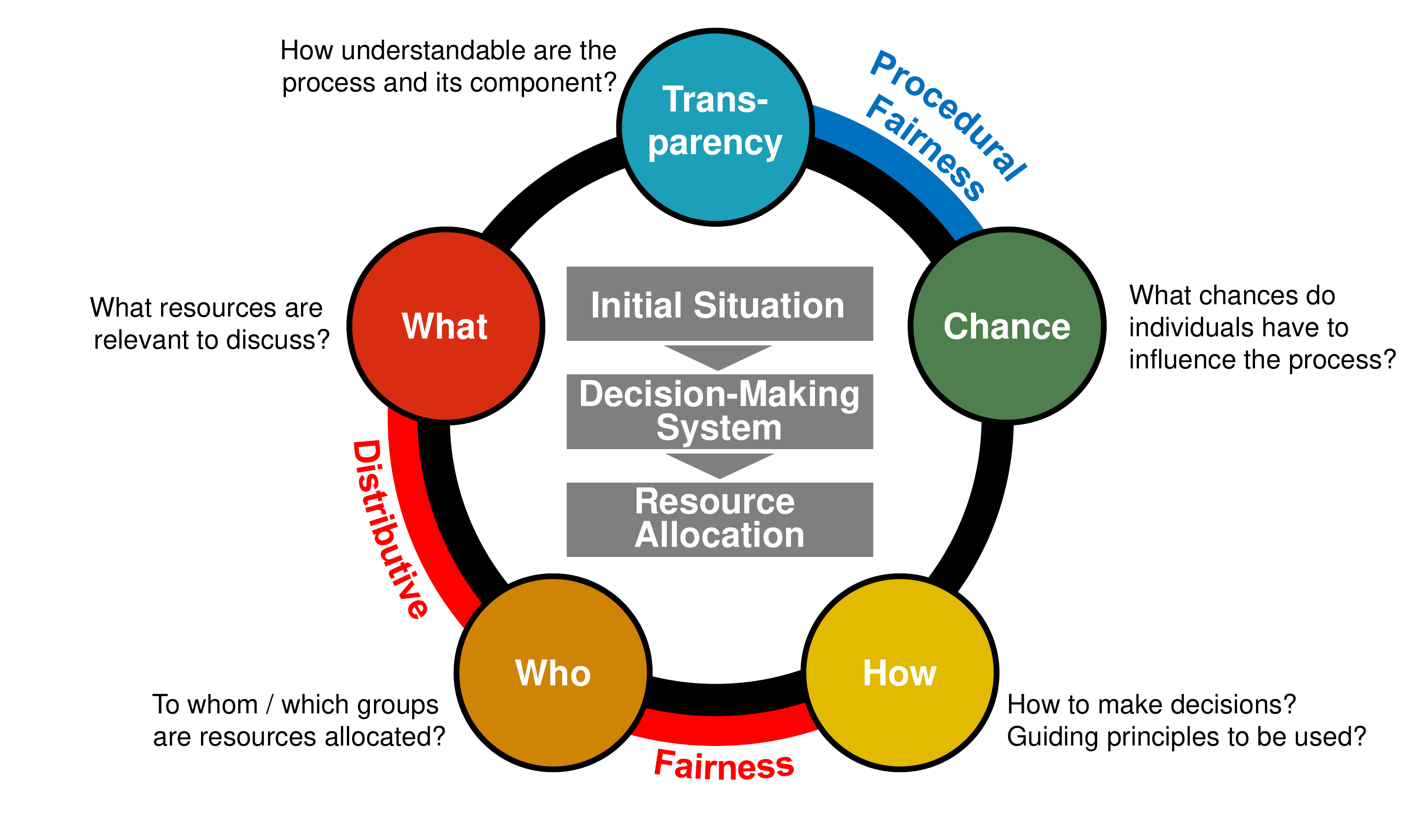}
    \caption{
        \textbf{The Fairness of Decision-Making Systems.} \\
        Decision-Making is depicted as a resource allocating process, that takes an initial situation as an input, makes a decision, and provides a resource allocation as an output. 
        The fairness discussion of decision-making therefore covers procedural and distributive fairness.
        For procedural fairness, aspects such as (i) the transparency of the process, and (ii) what chances individuals have to influence the process are important questions to answer.
        For distributive fairness, aspects such as (i) what resources are fairness-relevant and distributed to (ii) which groups and (iii) how decisions are made (based on which guiding principles) are important questions to answer. 
    }\label{fig1A}
\end{figure*}

Algorithm-driven, automated processes in some form distribute resources to people.
As these processes are automated by technology, a discussion of the fairness of cybernetic systems must be a discussion of procedural and distributive fairness~\citep{pereira2017distributive,friedman2019power} (Fig.~\ref{fig1A}).
For example, credit scoring algorithms determine whether a certain person has access to a loan, transportation demand management and congestion pricing determine whether a certain person has access to the road infrastructure of a city, automatic screening algorithms in recruiting determine whether a certain person is invited to a job interview, and user-engagement-maximizing algorithms determine which information is provided to the consumer of a social media platform.

While most discussions in the fairness literature are instrumental to discuss which resources can be considered as fairness-relevant~\citep{rawls1971atheory,walzer1983,nussbaum2011creating,sen2008idea}, which groups shall be compared and what fairness can be considered as conceptually~\citep{goppel2016handbuch}, only few answers can be found on how to quantitatively assess fairness in a specific situation based on data, which is crucial for the design and integration into algorithms, that in some form act within their environment using data~\citep{cormen2022introduction}.
The quantitative measures proposed in the literature are either too specific to a particular application, suffer from some undesirable characteristics, or are limited to specific ideologies~\citep{jain1984quantitative}.



It is the mission of this article (i) to highlight the importance of a quantitative discussion on the fairness of cybernetic systems, (ii) to enable algorithm design and evaluation based on a holistic, ideology-agnostic, quantitative fairness framework for a distributive discussion, and (iii) to create a connection between domain-specific literature and the fairness literature.

The remainder of this work is organized as follows.
Section~\ref{review} reviews the existing literature of fairness, distributive justice, domain-specific discussions of fairness in various fields, and elaborates in particular on algorithmic fairness, biased training data, and the challenges of algorithm transparency and explainable artificial intelligence (AI).
Section~\ref{method} proposes the quantitative fairness framework for distributive justice.
Section~\ref{discussion} discusses the usefulness of the proposed fairness framework and illustrates its application when designing algorithms for automated decision-making.
Section~\ref{conclusion} concludes this work.

\section{Related Works} \label{review}

Exploring the concept of fairness presents significant challenges, as it is an abstract philosophical notion deeply rooted in specific social and cultural contexts. 
Despite over two thousand years of philosophical discourse across diverse human civilizations, a universal consensus on the definition and application of fairness remains elusive.
Fairness is widely regarded as a cornerstone of human coexistence and is extensively examined as a multidisciplinary concept across various fields, including philosophy, ethics, biology, sociology, political science, economics, and religion studies.
The terms fairness, equity and justice are used interchangeably in the literature, where equity is often found in the literature of economics and justice in the literature of politics and law~\citep{goppel2016handbuch}.

In this section, we start to discourse distributive fairness and procedural fairness.
Then, we review domain-specific discussions of fairness to capture its multidisciplinarity.
As automated decision-making impacts almost every aspect of our life, a review of various domains enables us to gather a broader perspective for the development of a quantitative fairness framework for resource allocation in equitable, cybernetic societies.
At the end, we elaborate on the algorithm aspects of cybernetic societies: algorithmic fairness related to biased training data, algorithmic transparency, and explainable AI.

\begin{figure}[ht!]
    \centering
    \includegraphics[width=0.45\textwidth]{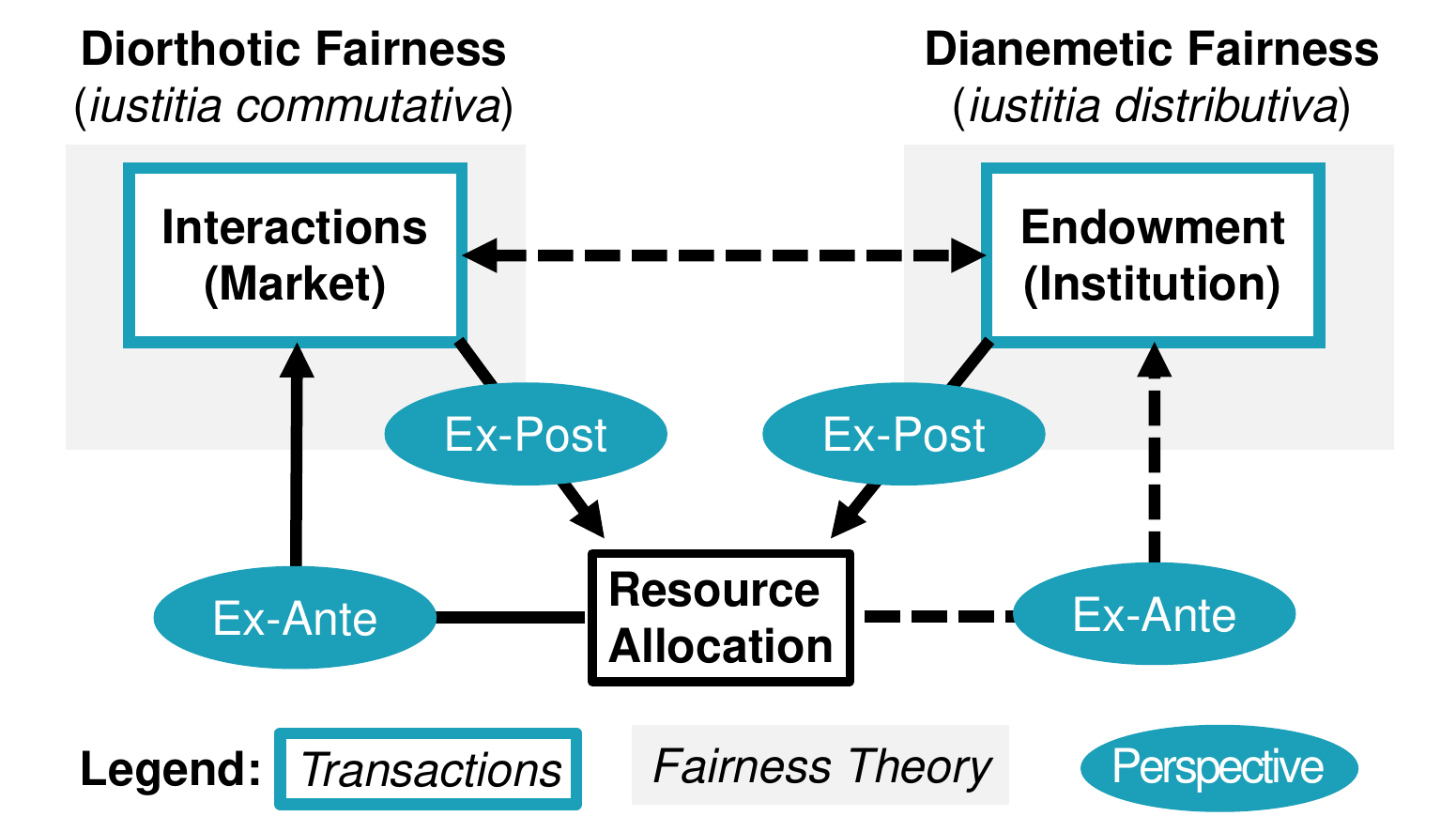}
    \caption{
        \textbf{A Transactional View on Distributive Fairness.}
        Aristotle distinguishes dianemetic and diorthotic, distributive fairness. Dianemetic fairness is concerned with the fair distribution of resources from top to bottom (endownment) from usually one central decision maker (institution, government) to the population. Diorthotic fairness is concerned with the fair distribution of resources in a decentralized way as a result of transactions between individuals (markets).
    }\label{fig_transact_framework}
\end{figure}

\subsection{Distributive Fairness}
Distributive fairness  (also distributive justice) tries to find fair ways of allocating resources (goods, opportunities, etc.) across a population as a result of transactions. This can include how benefits but also burdens can be distributed across individuals, reflecting or neglecting individual properties such as wealth, social status, contributions, needs, etc.
\cite{hume1739} requires three conditions to discuss the question of fairness: scarcity of resources, a conflict of interest, and a relative balance of power between the negotiating parties.
Important aspects which need clarification in the discussion of distributive fairness include: (i) which resource is distributed (ii) across whom and (iii) how (based on which guiding principle).

Three important concepts are useful to determine the fairness-relevance of a specific resource include: primary goods, distributive spheres of justice, and capabilities. 
\cite{rawls1971atheory} contends that primary goods, which are universally valuable and impact the well-being of all individuals regardless of their personal preferences, are crucial considerations in determining fairness.
\cite{walzer1983} argues that goods with a distinct social meaning need to be distributed in dedicated, fair, distributive spheres, contrary to normal goods. 
The capability approach advocates to discuss resources that determine capabilities (freedom of choice through many opportunities) rather than functionings (actual outcomes)~\citep{nussbaum2011creating,sen2008idea}.

\begin{figure}[ht!]
    \centering
    \includegraphics[width=0.45\textwidth]{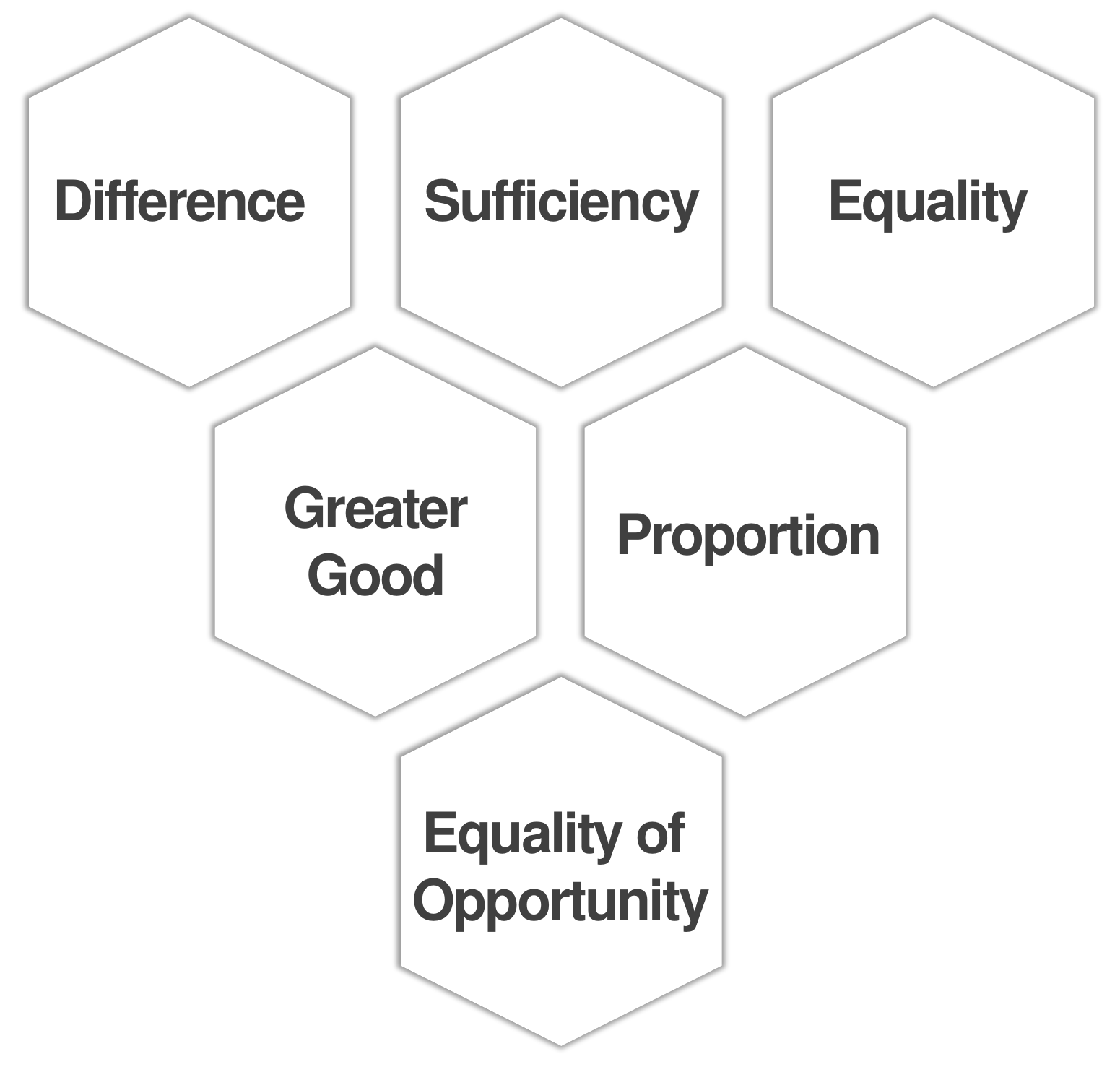}
    \caption{
        \textbf{Philosophic Guiding Principles For Fairness.} \\
        Six guiding principles for distributive fairness that cover different perspectives on fairness.
        The difference principle advocates a situation in which the least-advantaged (poorest) are in the best possible situation.
        The equality principle advocates an equal distribution of outcomes.
        The equality-of-opportunity principle advocates an equal distribution of the initial situation to provide every individual with the same chances to make their own luck. 
        The greater-good principle is the basis for Utilitarianism and argues that the suffering of the few is acceptable if it serves the greater good of the majority. 
        The proportion principle advocates distributions that stand in proportional relation to the contributions and status of individuals.
        The sufficiency principle advocates distributions in which everyone receives a certain sufficient minimum to satisfy basic needs, inequalities above that are not an issue.
    }\label{fig1B}
\end{figure}

Aristotle's Nicomachean Ethics~\citep{wolf2002aristoteles} presents a theory of transactional fairness that distinguishes between two types of justice: dianemetic and diorthotic fairness (Fig.~\ref{fig_transact_framework}). 
Transactions can occur between various parties under different circumstances, including market exchanges, governmental resource allocation, participation in political decision-making, or legal proceedings. 
Dianemetic fairness focuses on how resources are allocated from an authority to a population, typically involving governmental institutions in cases such as subsidies or societal redistribution. 
The initial distribution of resources can also be considered a dianemetic endowment. 
Diorthotic fairness, on the other hand, deals with transactions between individuals or groups within a population, often in the context of market exchanges. 
\cite{nozick1974anarchy} bridges the gap between these two concepts by arguing that fair, diorthotic transactions can only occur if the initial, dianemetic allocation was already fair. 

\cite{goppel2016handbuch} distinguish two types of perspectives on fairness: ex-post, and ex-ante fairness.
These perspectives describe what matters for the fairness of a resource allocation; the first one (ex-post) focuses
only on the output of a transaction, the second (ex-ante) one focuses only on the input of a transaction. 
Teleological fairness focuses on outcomes or consequences, while deontological fairness focuses on the adherence to rules, duties, and principles regardless of the outcome. 
Moreover, a differentiation between horizontal equity (fairness within groups of similar individuals) and vertical equity (fairness between different groups) is possible.

The literature of fairness provides various ideologies and moral guiding principles, that can be summarized by following six principles: equality, proportion, greater-good, difference, equality-of-opportunity, and sufficiency (Fig.~\ref{fig1B}).
The equality principle considers an allocation fair, if it ensures equal allocation for all. This principle is the foundation for Egalitarian ideology~\citep{scheffler2017egalitarianism}.
The proportion principle considers an allocation fair, if allocated resources stand in proportion to the status of individuals, where status could refer to social status in a society, economic power, contributions to the system, or needs. This principle is the foundation for Aristocratic (also Aristotelian) ideology~\citep{goppel2016handbuch,wolf2002aristoteles}.
The greater-good principle considers an allocation fair, if it maximizes the welfare of the many (society), even if this means the suffering of few (individuals). This principle is the foundation for Utilitarian ideology~\citep{mill2016utilitarianism}.
The difference principle considers an allocation fair, if it achieves the best possible outcome for the least-advantaged (Rawlsian ideology) or the average (Harsanyian ideology)~\citep{rawls1971atheory,harsanyi1975can}.
The equality-of-opportunity principle considers an allocation fair, if the initial changes / opportunities of each participating individual to a resource-allocating process were equal upfront. This principle is the foundation for Luck-Egalitarian ideology~\citep{dworkknork}.
The sufficiency principle considers an allocation fair, if it is guaranteed that each individual receives at least a sufficient minimum. This principle is the foundation for Sufficientarian ideology~\citep{shields2012prospects}.


\subsection{Procedural Fairness}
The concept of fairness has evolved throughout history. In ancient philosophy, it was viewed as a personal virtue and a key attribute of divine beings. During the Renaissance, this divine association shifted towards a universal, natural law. The Enlightenment era saw philosophical discussions move away from religious contexts, emphasizing rational thought instead. Contemporary debates on fairness primarily center on political and economic aspects, expanding beyond individual conduct to encompass institutional practices and societal processes~\citep{goppel2016handbuch}.

Procedural fairness discusses fairness in processes.
These processes usually include processes that resolve disputes and distribute resources, particularly in administrative and legal contexts, but are not limited to those. 
Other examples could include promotions to higher positions in organizations, hiring for new positions, admission to educational facilities, etc.
Procedural fairness intersects with distributive fairness (when distributing resources), and retributive fairness (when punishing mistakes).
Two important aspect to procedural fairness are (i) the opportunities of users affecting the process outcome, and (ii) the transparency of the processes and underlying decision-making~\citep{lind2013social}.
Similarly to distributive fairness, ex-ante and ex-post perspectives on procedural fairness apply.

Three important schools of thought exist in the equality-of-opportunities of processes: formal equality, substantive equality, and pure equality.
Formal equality (also meritocracy) often refers to meritocratic provision of opportunities based on performance only, neglecting other discriminatory aspects such as gender, race, or age.  
Substantive equality refers to equal chances for different groups of people. 
For example, substantive equality distributes opportunities according to gender and race, rather than merits, to ensure the representation of certain groups.
While formal and substantive equality discriminate individuals either based on performances or personal aspects (e.g. gender, race), the pure equality (also known as equality before the law) describes the equal distribution of opportunities to all individuals independent of any personal features~\citep{acemoglu2021theory,barnard2000substantive}.

\subsection{Domain-specific discussions of fairness}

The discussions of fairness across different domains have in common, that fairness - besides efficiency - is a crucial component when designing systems.
Not only does fairness promote cooperation between selfish individuals to achieve better outcomes at the system (population) level, but also to confront public resistance against policy, to gain trust and acceptance for rules and institutions, to perforate self-reinforcing cycles of poverty through social mobility, to incentivize motivation, contribution and satisfaction of people through inclusion, to increase social-cohesion in groups and group performance, and ultimately to improve life quality. 

\subsubsection{Natural Science \& Engineering}
\textbf{Biology \& Behavioural Psychology - }
Fairness in biological systems often manifests as cooperative behaviors. 
From an evolutionary biology perspective, sensitivity to fairness is a behavioral trait that evolved in social animals, whose survival relies on cooperation and group dynamics. 
This perception of fairness can be observed in various social animals, including Capuchin monkeys, vampire bats, and humans~\citep{brosnan2014evolution}.
At the individual level, the subjective perception of fairness plays a crucial role in both physical health and psychological well-being. 
The brain of social animals, particularly the insula, reacts to perceived unfairness with a sense of disgust. 
Interestingly, this reaction is more pronounced when individuals feel they are unfairly disadvantaged compared to when they are unfairly advantaged~\citep{jackson2006linking}. 
Furthermore, perceptions of fairness significantly influence how individuals form relationships and interact with one another~\citep{de2010procedural}.
At the societal level, fairness plays a crucial role in fostering robust communities, encouraging adherence to social norms, and promoting cooperation. 
The perception of fairness is shaped by shared, group-specific norms and cultural values. Moreover, fairness serves to enhance social cohesion, mitigate conflicts, bolster group identity and trust, and ultimately leads to improved group work outcomes~\citep{de2010procedural,hitti2011social}.
The evolution of fairness in biological systems is closely tied to the concept of reciprocal altruism. Organisms that engage in fair exchanges are more likely to maintain beneficial relationships over time, increasing their chances of survival and reproduction. 
This has led to the development of various strategies and mechanisms in nature to enforce fairness, such as partner choice and reputation systems in some species. 
Often, fairness reflects a balance between contribution and reward that is sustainable and beneficial for the species or ecosystem as a whole~\citep{schino2009reciprocal}. 

\textbf{Network Management - }
Human societies use networks for the transportation and distribution of resources, e.g. energy networks, telecommunication networks, supply chain networks, rail- and road networks, computation networks, and social networks.
Fairness in network management is a critical aspect of ensuring that resources are distributed equitably among users and applications. This principle is essential for maintaining the efficiency, reliability, user satisfaction, and motivation to contribute to a network.
Discussions of fairness in networks include allocated resources, but also quality of service (e.g. delays or bandwidth in the context of internet networks). 
Achieving fairness in network management often involves trade-offs between efficiency and equity. For instance, while it is desirable to allocate resources fairly, it is also important to ensure that the network operates efficiently and maximizes overall throughput. This balance can be challenging, particularly in heterogeneous networks where users have varying requirements and capabilities. Utility optimization methods are often used to address this issue, where the goal is to maximize the aggregate utility of all users while ensuring fair resource allocation.
Various fairness models, such as max-min fairness, proportional fairness, $\alpha$-fairness, and the Jain metric have been developed to address these allocation challenges~\citep{jain1984quantitative,bonald2006queueing}.

\subsubsection{Economics}

\textbf{Macro Economics - }
Economics studies the production, consumption and distribution (allocation) of resources in human societies.
Fairness and economics are deeply intertwined concepts that have significant implications for market structures, competition, policies and taxation.
Governmental intervention into markets is often justified with restoring competition that is affected by monopolies, public goods, and externalities.
Minimum wages, wealth redistribution and income taxation play an important role in fairness-promoting policies.
Behavioral economics has provided new insights into how people form judgments about what is fair. 
Factors like framing, social norms, and perceptions of intentions all influence whether a given outcome is seen as fair or unfair, which has important implications on public support for policies.
Welfare economics is a branch of economics that studies social welfare, and to evaluate the overall well-being of a society, where fairness is an important component to most definitions of welfare. Especially welfare-economists have shaped the fairness literature of the recent. 

\textbf{Micro Economics \& fairness of market prices - }
When can markets and market prices considered as fair?
The fairness of markets was closely linked with the philosophical, political and economic discussions of fairness.
Aristotle argues, a diorthotic transaction is fair, when the exchanged resources are of equal value.
Albertus Magnus and Thomas Aquinas introduced the term of a fair price for transactions at monetary markets. 
The fair price primarily reflects the efforts for the generation of the resource, but can also include marginal profits of traders.
The school of Salamanca puts the term of a fair price equal to the market price, assuming efficient, ideal markets.
\cite{smith1937wealth} developed the theory of the invisible hand which claims, that any selfish, egoistic behavior and any price in transactions is fair, as free markets lead to societal optima as a result. 
Besides market failure and arbitrage, purposeful phenomena such as price differentiation, dynamic pricing, price discrimination, and personalized pricing are heavily discussed in the literature. 
The fairness of markets is therefore closely linked to the fairness of prices~\citep{kahneman1986fairness}.

\textbf{Education - }
Education systems are a main ingredient for societal development and individual growth.
Education is a catalyst and one of the most powerful tools for promoting social mobility.
It provides individuals with knowledge, skills, and opportunities that can lead to better job prospects and higher earning potential, potentially allowing them to improve their socioeconomic status~\citep{brown2017education}.
Especially early childhood education has been identified as a key strategy for promoting social mobility.
Fairness plays an important role in this context to make sure, that each child has the chance to grow up to its full potential.
Education offers resources and opportunities.
As children have different socio-economic backgrounds, abilities, development stages, and abilities when entering the school, access to the same quality of teachers, textbooks, and learning environments are important.
However, in addition to that, more resources and opportunities might be necessary for students with special needs, such as inclusion of children with disabilities.
Governments play a crucial role in providing public education systems, to overcome the significant disparities in society that hinder children from disadvantaged contexts~\citep{tharp2018teaching}. 

\textbf{Housing \& gentrification - }
Gentrification is a complex urban phenomenon that sits at the intersection of housing, economic development, and social equity. 
While it can bring positive changes to neighborhoods, such as improved infrastructure and increased economic activity, it also raises significant concerns about fairness and displacement of long-time residents.
At its core, gentrification involves the influx of wealthier residents into previously lower-income neighborhoods, often accompanied by rising property values and rents, and changes in the local culture and amenities. 
This process can lead to the displacement of long-term residents who can no longer afford to live in their communities, which affects social cohesion, and raises important questions about housing fairness and social justice~\citep{krings2020equity}.
In many societies, the housing market and rental prices are therefore highly regulated, and challenges to account for discrimination, racial segregation, economic disparities, and marginalised communities bring complexity to this issue~\citep{von2000study}.

\textbf{Healthcare - }
Access to medical services is vital for sustaining healthy and worth-living lives, and for keeping up productivity of the workforce. 
In many societies, access to healthcare and quality of treatment outcomes depends upon socio-economic contexts.
At the same time, disease and sickness have been identified as one of the most important reasons for people turning into poverty (and even homelessness in extreme cases)~\citep{jamison2018disease}.
Fairness-relevant discussions in this context include health insurance systems on the societal level, and various ethical questions on the individual level, such as prioritization of patients in the context of emergency rooms and organ transplants~\citep{daniels1996benchmarks,ding2019patient}.

\textbf{Transportation planning \& policy - }
Transportation infrastructure design determines how effectively people and goods can move around.
Fairness arises as an important theme in the field of transportation.
Accessible, affordable, safe, inclusive, and barrier-free transportation are the key aspects of discussion.
The planning of transportation infrastructure, such as roads, railways, and public transport are important to enable an equitable access for as many as possible, with positive outcomes for the economy and life-quality.
Often, transportation infrastructure faces a demand which is higher than its supply, resulting in congestion, and long waiting queues; policies for traffic demand management such as congestion pricing are part of highly controversial debates. 
Especially fairness and equity-concerns are the major impediments for the real-world implementation of transportation policies~\citep{martens2016transport,gu2018congestion}.

\subsubsection{Business administration}

\textbf{Management - }
Fairness in management is a critical component of effective leadership and organizational success. It encompasses the equitable treatment of employees, transparent decision-making processes, and the creation of inclusive work environments. When managers prioritize fairness, they foster trust, boost morale, and enhance overall productivity within their teams.
Consistent and unbiased treatment of employees, transparency in decision-making processes, equal opportunities for growth and development, conflict resolution, and compensation and recognition practices count amongst the most important fairness-relevant aspects of management.
By prioritizing equitable treatment, transparency, equal opportunities, and impartial conflict resolution, managers can create a work environment where all employees feel valued and respected. 
This approach not only benefits individual team members, but also contributes to the long-term success and sustainability of the organization as a whole~\citep{simons2003managers}.

\textbf{Recruiting \& hiring processes - }
Ensuring fairness in recruitment and selection not only promotes diversity and inclusion, but also enhances the reputation of the organization as an equitable employer, increases the chance to find the most skilled workers for a position, and improves organizational efficiency.
Procedural fairness, transparency on the recruiting process, and bias awareness play an important role.
To account for diversity and inclusion, both the applicant pool diversity and hiring outcome diversity are common measures to quantify fairness in this context.
The increased use of automated decision-making allow for the consideration of larger number of applications, but also harbor the threat for technological biases, that need to be considered carefully~\citep{van2020hiring}.

\textbf{Banking \& loan approval - }
Access to financial services, regardless of race, gender, residence, and other factors, is important to achieve inclusion and participation of customers.
Historically, certain demographic groups, particularly those from marginalized communities, have faced systemic barriers, especially in obtaining loans.
These barriers can include discriminatory lending practices, biased credit scoring models, and a lack of transparency in loan approval processes. 
Addressing these issues is crucial for creating a more equitable financial landscape and a greater trust of the public in the industry.
With the increasing use of machine learning and algorithms in credit scoring and loan approval, the concept of algorithmic fairness has gained prominence. Algorithms can inadvertently perpetuate biases present in historical data, leading to unfair outcomes for certain groups. 
Therefore, various fairness metrics and bias mitigation strategies have been proposed to ensure that algorithms do not discriminate against marginalized groups. 
While fairness is crucial, financial institutions must also consider risk management in their lending practices. Banks need to assess the creditworthiness of applicants to minimize default risks. 
However, this assessment should be conducted in a way that does not disproportionately impact certain groups. 
Striking a balance between ensuring fair access to loans and managing financial risk is a complex challenge that requires innovative approaches and ongoing evaluation~\citep{lee2021algorithmic}.

\subsubsection{Social Sciences}

\textbf{Social justice - }
Social justice is a concept that encompasses the fair and equitable distribution of resources, opportunities, and privileges within a society. It is rooted in the principles of equality, human rights, and collective responsibility. 
The pursuit of social justice aims to address systemic inequalities and promote a more inclusive and just society for all individuals, regardless of their background or circumstances. 
At its core, social justice seeks to rectify historical, ongoing, and self-reinforcing disparities in areas such as economic inequality (e.g. rich vs. poor), racial, and ethnic inequalities (e.g. white vs. black), gender equality (e.g. men vs. women), or disability rights.
The pursuit of social justice often involves challenging existing power structures and advocating for systemic changes. 
This can include policy reforms, grassroots activism, education, and awareness campaigns. Social justice movements have played a crucial role in advancing civil rights, workers' rights, and other progressive causes throughout history~\citep{harvey2010social}.
The digital age has brought new dimensions to social justice efforts, with social media and online platforms serving as powerful tools for organizing, raising awareness, and amplifying marginalized voices. However, it has also highlighted new challenges, such as digital divides and online harassment~\citep{eubanks2012digital}.

\textbf{Environmental \& intergenerational justice - }
A particular form of social justice discusses the interplay between humans and their ecologic environment, as well as how social justice between generations in the presence of demographic transformation and climate change can be addressed. 
Environmental justice focuses on ensuring that all people, regardless of race, color, national origin, or income, have equal protection from environmental and health hazards. It emerged as a social movement in the United States in the 1980s, highlighting how marginalized communities often bear a disproportionate share of environmental risks. This movement has since expanded globally, addressing issues such as hazardous waste disposal, resource extraction, and land use that negatively impacts vulnerable populations~\citep{mohai2009environmental}.
Intergenerational justice extends this concept across time, emphasizing the responsibility of current generations to preserve the environment for future ones. It recognizes that today's actions have long-term consequences that will affect the quality of life and opportunities available to future generations~\citep{barry1997sustainability}.

\textbf{Sports \& competitions - }
Fairness is a fundamental principle in sports and competitions, serving as the foundation for meaningful and equitable contests. 
Equal opportunity in the sense of skill, effort, and merit rather than external advantages, is at the core of fairness in sports. 
Sport organizations strive to create rules and structures that give competitors an equal chance to succeed based on their abilities and preparation.
Fairness also involves adherence to established rules and regulations. These rules define the boundaries of acceptable behavior and performance, ensuring that all participants compete under the same conditions. Enforcement of rules, including anti-doping measures, helps maintain the integrity of competitions and prevents unfair advantages.
The principle of fair play extends beyond just following rules. It values sportsmanship, respect for opponents, and ethical behavior both on and off the field. This broader concept of fairness contributes to the overall positive culture and values associated with sports.
Fairness in sports also has implications beyond individual competitions. It plays a crucial role in maintaining public trust and interest in sports. When spectators believe competitions are fair, it enhances the excitement and credibility of the events, contributing to the overall appeal and sustainability of sports~\citep{loland2010fairness}.

\subsubsection{Government \& policy}

\textbf{Policy making -}
Fairness is a critical consideration in policy development and implementation across various domains of governance and public administration.
Procedural fairness is essential in policy creation and enforcement. Policies should be developed through transparent processes that allow for public input and stakeholder consultation. When implementing policies, authorities must apply rules consistently and provide clear explanations for decisions. 
Policymakers must consider how different groups are impacted and strive for outcomes that are perceived as just. This may involve targeted interventions to address historical inequities or support vulnerable populations.
Policies should aim to promote equality-of-opportunity while recognizing that strict equality of outcomes is not always achievable or desirable. Anti-discrimination policies, for example, seek to ensure fair access and treatment across gender, racial, and other lines.
Evidence-based policy-making can help promote fairness by relying on objective data rather than subjective biases. 
Perceptions of fairness can vary across cultural contexts and change over time. 
Policymakers must remain attuned to evolving societal values and engage in ongoing dialogue with diverse communities~\citep{gilley2017technocracy}.

\textbf{Legal system \& criminal justice -}
Fairness in the legal system is a principle that underpins the rule of law and ensures that justice is administered equitably and impartially. 
This principle is crucial for maintaining public trust and confidence in the legal system.
Procedural fairness in this context includes the right to a fair hearing, and the right to an impartial decision-maker. 
In criminal cases, procedural fairness is essential to protect the rights of the accused, including the presumption of innocence until proven guilty, the right to legal representation, and the right to a fair trial. 
Ensuring that these procedures are followed helps to prevent miscarriages of justice and upholds the integrity of the legal system.
Equality before the law is another aspect of fairness in the legal system. This principle asserts that all individuals, regardless of their background, are subject to the same laws and are entitled to equal protection under the law. 
This means that the legal system must be free from discrimination based on race, gender, ethnicity, socioeconomic status, or any other characteristic. 
Achieving true equality before the law requires ongoing efforts to address systemic biases and ensure that legal processes do not disproportionately disadvantage certain groups ~\citep{berk2021fairness,hurwitz2005explaining}.

\textbf{Police operations -}
Fairness plays an important role for policing.
Procedural fairness influences both the public's perception and acceptance of law enforcement, and the internal dynamics, efficiency and cohesion within police organizations.
When police officers engage in fair procedures, such as treating people with dignity, providing explanations for their actions, and being neutral in decision-making, they are more likely to be perceived as legitimate authorities. 
This perception of legitimacy is crucial because it fosters public trust and cooperation with law enforcement efforts, even when the outcomes are not favorable to the individuals involved.
A study involving New York City residents found that procedural fairness was a key antecedent of police legitimacy, which in turn influenced people's willingness to comply with the law~\citep{sunshine2003role}.
Racial profiling in law enforcement is a complex and contentious issue that highlights the tension between fairness and perceived effectiveness in policing.
At its core, racial profiling uses race or ethnicity as a key factor in deciding whether to stop, search, or investigate individuals. 
While proponents argue it can be an efficient crime prevention tool, racial profiling raises serious concerns about discrimination, civil rights violations, and the erosion of public trust in law enforcement.
Racial profiling is controversial because it judges individuals based on group characteristics rather than individual behavior or evidence. 
This can lead to a self-fulfilling prophecy, where increased investigation of certain groups results in higher findings and arrest rates, which are then used to justify further profiling~\citep{hurwitz2005explaining}.

\subsection{Algorithmic Fairness}

Algorithms can be used to formalize and automate decision-making processes.
An algorithm is a finite sequence of well-defined steps to process inputs and produce outputs.
There are two sources that can be used to define algorithms: First, algorithms can be formalized by humans, that specify inputs, outputs, and process-steps manually to mimic their own decision-making.
Second, the advances in computer science, especially in supervised machine learning, allowed computers to derive algorithms directly from data (input-output pairs).
Cybernetic systems use algorithms to control systems, and they cannot only be used by computers, but also by human-driven processes, such as governments that follow bureaucratic and legal processes, and by physical machines.

The fairness of algorithms massively depends on the choice of inputs, outputs, process-steps, and the source of the algorithm.
Algorithmic biases can lead to systematic and repeatable errors and discrimination of certain individuals or groups, and cause algorithmic unfairness.
Algorithmic unfairness in cybernetic systems can lead to self-reinforcing cycles of inequality, take racial profiling as an example: 

Algorithmic unfairness in cybernetic system, such as group discrimination that is even justified by data, can be harmful, as the Hirshleifer-effect~\citep{hirshleifer1978private} in the context of credit scoring illustrates: 
Assuming that discriminating a certain group when deciding on loan approvals is justified by data and increases efficiency of decisions (improved default-risk management), and as a consequence increases profits for the banks and enables lower loan fees for the other customers.
While using more data and develop better algorithms seems like a good idea on the individual firm level for banks and insurances, it might be harmful on a societal level.
The Hirshleifer-effect describes the paradoxical situation, that the release of more and better information to the public (market) mutually prevents beneficial risk-sharing trades, leading to a situation where all agents are worse off compared to a situation where fewer or incomplete information was available.

Algorithms with biases often stem from imbalanced data, where certain groups may be underrepresented or misrepresented, leading to unfair outcomes. Addressing these issues involves various strategies, including data preprocessing to balance datasets, in-process adjustments to algorithms during training, and post-process evaluations to ensure equitable outcomes~\citep{pessach2023algorithmic}.
The challenge of imbalanced data is particularly pronounced in fields like healthcare, where disparities in data collection can exacerbate existing inequalities in treatment and diagnosis. For example, the design and development of medical treatments have historically exhibited a gender bias, often optimizing for male physiology at the expense of female patients~\citep{agyemang2023uncovering}.



\subsection{Transparency and Explainable AI}

Increasingly, automated decision-making processes employ deep-learning with neural network models for machine learning, when developing automated decision-making algorithms.
These models have an ever-growing number of parameters and thus complexity, which allows them to achieve super-human performance (for example in image recognition and classification).
At the same time, these models are highly problematic, as due to their complexity it is challenging to understand how and based on which specific properties and features from the data, decision are made.
Transparency and traceability of decision-making processes are a crucial component to procedural fairness.
As a consequence, the application of even better performing models is impeded in industry- and governmental applications due to explainability concerns, the often unclear origin of training data and process, the inability to describe guarantees in terms of safety, and equity issues~\citep{xu2019explainable}.


\begin{table*} [ht!]
    \centering
    \begin{tabular}{llll}
        \hline
        \textbf{Principle} & \textbf{Perspective} & \textbf{Metric} & \textbf{Optimization} \\
        \hline
        Difference & Ex-Post & average or minimum of $y_i$ & maximize \\
        Equality & Ex-Post & dispersion of $y_i$ & minimize \\
        Equality-of-opportunity & Ex-Ante & dispersion of $x_i$ & minimize \\
        Greater-good & Ex-Post & sum of $u_i$ & maximize \\
        Proportion & Both & dispersion of ratio $y_i$/$x_i$ & minimize \\
        Sufficiency & Ex-Post & threshold share of $y_i$ & maximize \\
        \hline
        & & & \\
    \end{tabular}
    \caption{\textbf{Dianemetic Fairness Measures}. \\
    Statistical dispersion and economic concentration metrics are used to assess the equality of distributions. The formulas cover the most important measures and are denoted to determine the equality of a distribution of $n$ values $x_i$, where $\bar x$ represents the average over all $x_i$.}
    \label{tab:dianemetic_fairness}
\end{table*}

\begin{table*} [ht!]
    \centering
    \begin{tabular}{llll}
        \hline
        \textbf{Principle} & \textbf{Perspective} & \textbf{} & \textbf{Welfare Function} \\
        \hline
        Difference & Ex-Post &  & Rawlsian welfare function \\
        Equality & Ex-Post & & Sen and Foster welfare function \\
        Equality-of-opportunity & Ex-Ante & &  dispersion as Leontief-Lerner function \\
        Greater-good & Ex-Post & & Benthamite welfare function \\
        Proportion & Both & & none or dispersion of ratios $y_i$/$x_i$ \\
        Sufficiency & Ex-Post & & threshold-share of $y_i$\\
        \hline
        & & & \\
    \end{tabular}
    \caption{\textbf{Diorthotic Fairness Measures}. \\
    Statistical dispersion and economic concentration metrics are used to assess the equality of distributions. The formulas cover the most important measures and are denoted to determine the equality of a distribution of $n$ values $x_i$, where $\bar x$ represents the average over all $x_i$.}
    \label{tab:diorthotic_fairness}
\end{table*}

\begin{table*} [ht!]
    \centering
    \begin{tabular}{llll}
        \hline
        \textbf{Metric} & \textbf{Reference} & & \textbf{Formula} \\
        \hline
        
        & & & \\
        Atkinson-Index & ~\cite{atkinson1970measurement} & &
        $A_{\epsilon} = \left\{ 
                \begin{array}{l} 
                    1 - \frac{1}{\overline{x}} (\frac{1}{n} \sum_{i=1}^{n} x_i^{1-\epsilon})^{1/(1-\epsilon)}  \\ 
                    for \; 0 \leq \epsilon < 1 \\
                    \\ 
                    
                    1 - \frac{1}{\overline{x}} ( \Pi_{i=1}^{n} x_i )^{1/N}   \\ 
                    for \; 0 \leq \epsilon < 1 \\
                    \\ 

                    1 - \frac{1}{\overline{x}} \; \min_{i}(x_i)  \\
                    for \; \epsilon = + \infty \\
                    \\
                \end{array}
                \right. $ 
        \\
        
        & & & \\
        Gini-Coefficient & ~\cite{dorfman1979formula} & & $G = \frac{\sum_{i=1}^{n} \sum_{j=1}^{n} \| x_i - x_j \|}{2n \sum_{i=1}^{n} x_i}$ \\ 
        
        & & & \\
        Herfindahl-Index & ~\cite{herfindahl1997concentration} & & $HH_{n} = \frac{HH - \frac{1}{n}}{1 - \frac{1}{n}}, HH = \sum_{i=1}^{n} ( \frac{x_i}{\sum_{j=1}^{n} x_j } )^{2}$ \\
        
        & & & \\
        Hirschmann-Index & & & \\
        Hoover-Index& ~\cite{hoover1936measurement} & & $H = \frac{1}{2} \frac{\sum_{i}^{n} \| x_i - \overline{x}\|}{\sum_{i}^{n} x_i}$  \\ 
        
        & & & \\
        Palma-Index& ~\cite{palma2011homogeneous} & & $P = \frac{\int_{0\%}^{40\%} L(x)dx}{\int_{90\%}^{100\%} L(x)dx}$ \\
        
        & & & \\
        Standard Deviation & & & $\sigma = \frac{1}{n} \sum_{i=1}^{n} (x_i - \overline{x})^2$ \\ 
        
        & & & \\
        Theil-Index T& ~\cite{theilindex} & & $T_T = \frac{1}{n} \sum_{i=1}^{n} \frac{x_i}{\overline{x}} ln ( \frac{x_i}{\overline{x}})$ \\ 
        
        & & & \\
        Theil-Index L& ~\cite{theilindex} & & $T_L = \frac{1}{n} \sum_{i=1}^{n} ln ( \frac{\overline{x}}{x_i})$ \\ 
        
        & & & \\
        \hline
        & & & \\
    \end{tabular}
    \caption{\textbf{Dispersion metrics}. \\
    Statistical dispersion and economic concentration metrics are used to assess the equality and thus dispersion of distributions. The formulas cover the most commonly used measures and are denoted to determine the equality of a distribution of $n$ values $x_i$, where $\bar x$ represents the average over all $x_i$.}
    \label{tab:dispersion_metrics}
\end{table*}

\begin{table}[ht!]
    \centering
    \begin{tabular}{l|l}
        \textbf{Title} & \textbf{Formalization} \\
        \hline
         & \\
        \textbf{Leontief-Lerner} & $W = F(x_1,...,x_n)$ \\
         & \\
         & \\
        \textbf{Bergson-Samuelson} & $W = F(u_1,...,u_n)$\\
         & \\
        \quad \quad Isoelastic & $W = \frac{1}{1-\rho} \sum_{i=1}^{n} \alpha_i u_i ^{1-\rho}$\\
         & \\
        \quad \quad \quad \quad Benthamite & $W = \sum_{i=1}^{n} u_i $ \\
         & \\
        \quad \quad \quad \quad Rawlsian & $W = \min_{i=1}^{n} u_i $ \\
         & \\
        \quad \quad \quad \quad Bernoulli-Nash & $W = \Pi_{i=1}^{n} \alpha_i u_i$ \\
         & \\
         & \\
        \textbf{Capability-approach} & $W=F(y_1,...,y_n)$ \\
         & \\
        \quad \quad Sen & $W = \overline{y} ( 1 - G ) $ \\
         & \\
        \quad \quad Foster & $W = \overline{y} e^{-T_T}$ \\
         & \\
    \end{tabular}
    \caption{\textbf{Welfare functions}. \\
    Social welfare functions aim to define the social welfare $W$ of resource allocations, where $x$ describes the input to the resource allocation process (e.g. financial wealth), $y$ describes the output of the resource allocation process (e.g. allocated resources), and $u$ describes the utility of the allocated resources to the individuals. There are three types of welfare functions: Leontief-Lerner, Bergson-Samuelson, and Capability-approach welfare functions. }
    \label{tab:welfare_functions}
\end{table}

\section{Quantitative Fairness Framework} \label{method}

The quantitative fairness framework aims to provide a holistic tool-set to quantitatively assess the fairness of situations and hence make fairness accessible to algorithmic design in general.
This framework distinguishes itself from previous approaches, by its general applicability, integrative and ideology-agnostic approach, and its multi-perspective view including teleological and deontological considerations.
Rather than advocating a specific ideology, the framework enables an integrative analysis that combines different guiding principles and allows for systematic comparison.
In this section, we model decision-making processes as resource allocation mechanisms, and develop quantitative fairness metrics based on a transactional, distributive fairness discussion.

\subsection{Modelling as resource allocating process}
We model any decision-making system as a process with an input $x$ (initial situation), an output $y$ (resource allocation),
and differentiate the discussion of distributive fairness in the Nicomachean, transactional view as diorthotic and dianemetic fairness.
Arguably, any decision-making system in some form uses information about an initial situation to make an informed decision, and the outcome of any decision is related in some form to the distribution of resources, be it useful goods, opportunities to be selected, or burdens in retributive contexts.

While we value the importance of the transparency and chance aspect of procedural fairness in this context, we advocate that a distributive fairness discussion is more instrumental when designing algorithms for decision-making.
We acknowledge the challenge of a unifying definition and guiding principle for fairness, and therefore advocate the use of an integrative framework, that has the capacity to reflect multiple ideology, rather than to focus on dedicated ideologies. 

As a result, we provide a framework that is able to reflect following guiding principles: equality, proportion, greater-good, difference, equality-of-opportunity, and sufficiency. As not all of these ideologies combine the ex-post and ex-ante perspective on fairness, some will require input and / or output information. Additionally, following the Utilitarian ideology based on the greater-good principle, we denote the utility $u$ as a function $f$ of $y$: $u =f(y)$.
We consider a population with $n$ individuals, that each experience a treatment of the decision-making system, based on their individual input $x_i$, output $y_i$, and utility $u_i$.

In the following parts of this section we propose the use of statistic and dispersion measures for the quantification of dianemetic fairness, and the use of dedicated social welfare functions for the quantification of diorthotic fairness.

\subsection{Metrics for dianemetic fairness}

The suggested quantitative metrics for dianemetic fairness for different guiding-principles are summarized in Table~\ref{tab:dianemetic_fairness}. Following notes shall be mentioned in addition to Table~\ref{tab:dianemetic_fairness}:
\begin{itemize}
    \item As a note for the difference principle, the goal is the maximization of the average (Harsanyian) or minimum of any $y$ (Rawlsian).
    \item As a note for the sufficiency principle, the goal is the maximization of the share of all individuals for which the outcome exceeds a certain threshold $T$ ($y_i>T$).
    \item An Utilitarian interpretation of the difference, equality, proportion, and sufficiency principle could involve $u$ instead of $y$ as well. Similarly, the greater-good principle could also involve $y$ instead of $u$.
\end{itemize}

For equality, equality-of-opportunity, and proportion principle, the use of dispersion metrics is suggested.
We provide a compilation of common dispersion metrics in Table~\ref{tab:dispersion_metrics}. Some of them originate from statistics, where dispersion is a common term, while others originate from economics, where concentration is a common term.

\subsection{Metrics for diorthotic fairness}
We advocate the use of Welfare functions as quantitative metric for diorthotic fairness, as summarized in Table~\ref{tab:diorthotic_fairness} for different guiding-principles. 
Welfare functions are a formal representation of individual notions of welfare aggregated on a societal level.
Leontief-Lerner welfare functions describe welfare as a function of available resources.
Bergson-Samuelson welfare functions describe welfare as a function of individual utilities.
Capability-approach welfare functions, describe welfare as a function of incomes.

Table~\ref{tab:welfare_functions} shows different formalizations of the three presented groups of welfare functions.
The isoelastic welfare function is a general formalization of the Bergson-Samuelson welfare function, with the possibility to weight each individuals utility with $\alpha_i$; for $\rho=0$ it becomes the Benthamite welfare function in the sense of utilitarian fairness; for $\rho=+\infty$ it becomes the Rawlsian welfare function in the sense of the Rawlsian fairness; for $\rho \to 1$  it becomes the Bernoulli-Nash welfare function; for any other $\rho$ it is an intermediate welfare function between the extremes of utilitarian and Rawlsian fairness.
Sen and Foster apply dispersion metrics such as the Gini-Coefficient $G$ and the Theil-Index $T_T$.

Following notes shall be mentioned in addition to Table~\ref{tab:diorthotic_fairness}:
\begin{itemize}
    \item The Harsanyian ideology based on the difference principle could be reflected by the average of $x_i$ as Bergson-Samuelson Welfare function.
    \item The Sen and Foster Welfare function are not only concerned with equality, but also with the average $y_i$. In this context, the welfare functions not only capture relative distribution but also absolute distribution of resources. For example, this implies that if a slightly more unequal distribution of resources could increase the average resource received, this would be preferred over perfect equality. This could make sense in the context of resources, that deteriorate by splitting them into many small pieces.
    \item For the equality-of-opportunity principle could employ dispersion metrics as Leontief-Lerner functions (based on the inputs $x_i$. Contrary to the equality principle, only the relative distribution would play a role here, as the decision-making process is assumed to affect the output only, but not to affect the input.
    \item For the proportion principle one could use the dispersion of ratios as Welfare function. The proclaimer of this principle, Aristotle, argued that any market activity and resulting prices and outcomes are fair per se, if these are the result of free decisions of individuals. 
    \item For the sufficiency principle one could use the threshold-share of $y_i$ as Welfare function.
\end{itemize}

\section{Discussion} \label{discussion}

In this section we want to showcase the usage of the proposed, quantitative fairness framework at two exemplary case studies.
The fair cake-cutting problem is used to illustrate dianemetic fairness.
The daily work of two fishermen is used to illustrate diorthotic fairness.
Afterwards, we discuss the benefits of the proposed framework in terms of holistic, ideology-agnostic, transitive, and quantitative discussions, and provide some further notes on the complexities of the equality-of-opportunities guiding principle.

\begin{figure*}[ht!]
    \centering
    \includegraphics[width=0.75\textwidth]{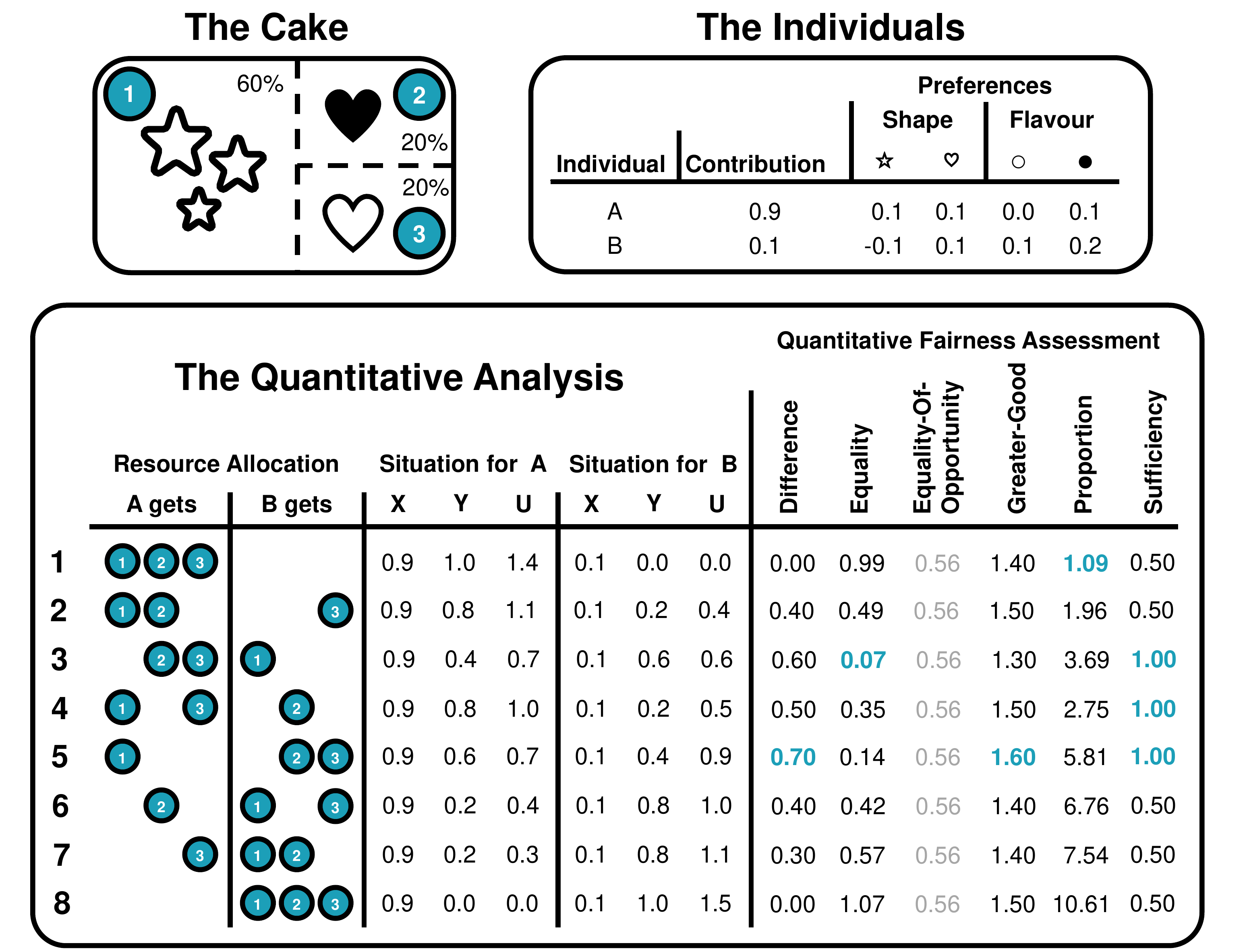}
    \caption{
        \textbf{Case Study: Fair cake-cutting problem \& dianemetic fairness.} \\
        A given cake that can be cut into three heterogeneous pieces only shall be distributed across two individuals, that differ in their contributions to paying or making the cake, and their preferences on the toppings. Eight allocations of the three pieces to the two agents are possible. Different guiding principles on fairness result in different, fairness-optimal recommendations for how to distribute the cakes.
    }\label{fig_casestudy_cakecut}
\end{figure*}

\begin{figure*}[ht!]
    \centering
    \includegraphics[width=0.75\textwidth]{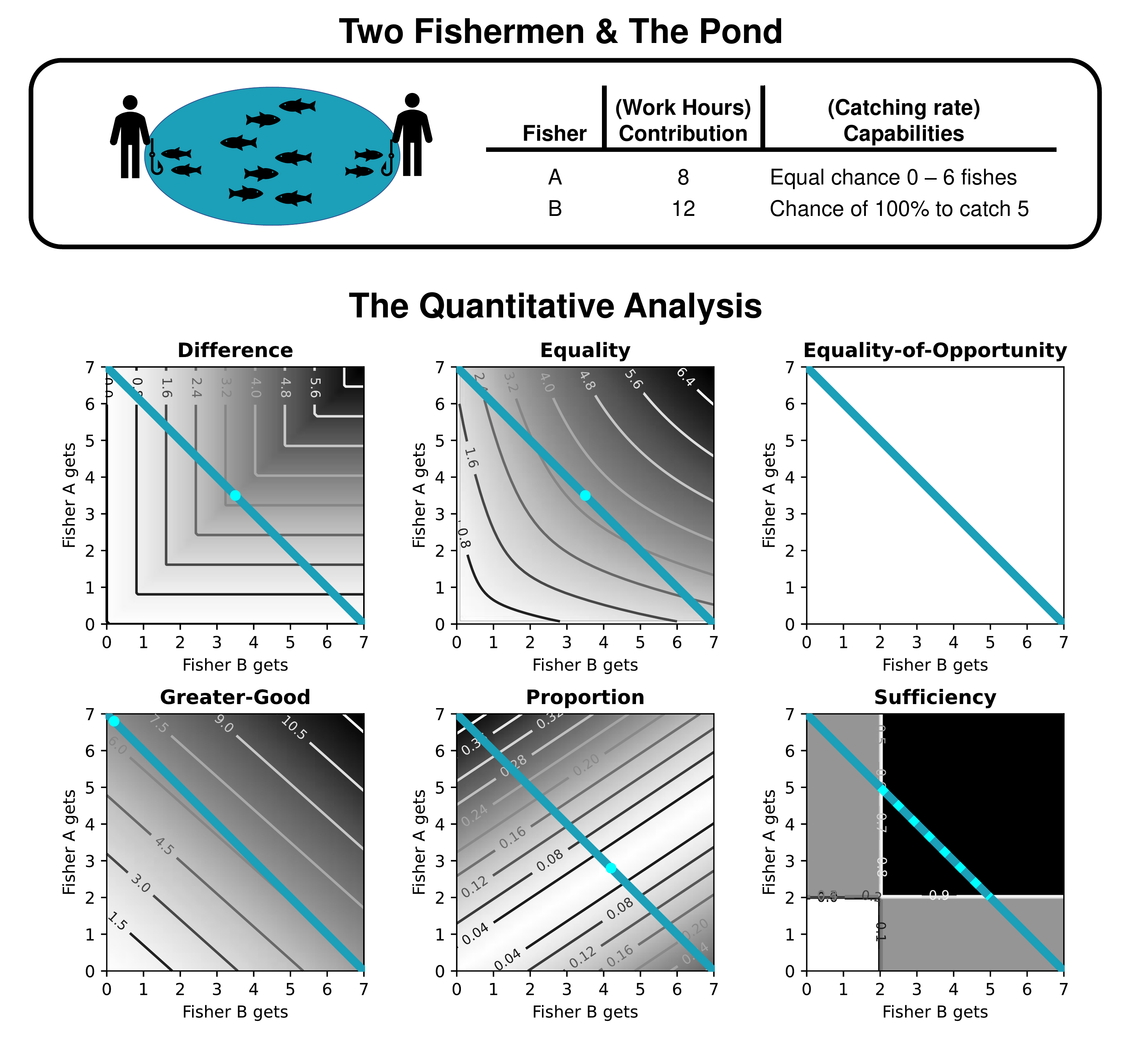}
    \caption{
        \textbf{Case Study: Fishermen \& diorthotic fairness.} \\
        Two fishermen go fishing at a pond every day, and at this specific day they fish seven fishes together. They differ in the working hours and catching rates due to different fishing techniques. Their outcome is stochastic. How should they distribute the total fish catch per day? Assuming fish is divisible, there is a continuous allocation space of the fish ranging from zero to seven fishes for each fisher. The Pareto-efficient frontier (blue line) displays this solution space. Different guiding principles on fairness result in different heatmaps and social welfare functions (contour plots of the heatmaps), with different recommendations on how to distribute the daily fish catch.
    }\label{fig_casestudy_fishing}
\end{figure*}

\subsection{Cake-cutting problem \& dianemetic fairness}
The fair cake-cutting~\citep{brams1996fair} is a typical division problem discussed in economics and dianemetic, distributive justice. There are different types of problems, including whether only the size of the cake piece matters (homogeneous goods) or also other features such as the toppings (heterogeneous goods), whether the cake can be cut everywhere (fully-divisible) or only to specific discrete pieces (partially-divisible), and whether the times of cutting creates waste of no use for anyone (non-lossy vs. lossy division).

The fair cake-cutting problem is a metaphor for various dianemetic, distributive fairness contexts: 
\begin{itemize}
    \item Imagine a manager that must distribute projects to staff; staff members can usually be staffed fully on one project only, and personal preferences and interests will affect how much staff members can learn from or are willing to invest into a project. Usually, it is the manager's decision, and therefore a dianametic context. This could be an example for a heterogeneous, partially-divisible cake-cutting problem.
    \item Imagine a government that tries to stimulate its stagnating economy during a recession with subsidies, tax-reductions, or cash subsidies amongst its population. How should different parts of the population (or companies) benefit from this support programme, e.g. it could make sense to support family households more than single households. This could be an example for a homogeneous, fully-divisible cake-cutting problem.
    \item Imagine a traffic signal controller at an intersection. It distributes delays to movement phases (respectively green time for passage). Every time there is a transition from one green movement phase to another, there is a short amount of time where both phases are red for security. Transitioning is important so that queues and waiting times do not get too long, but too frequent transitions will cause wasted time, where no vehicle can pass the intersection. How should the green time be divided to different movement phases? This could be an example for a lossy, fully-divisible cake-cutting problem.
\end{itemize}

For this case study, let us consider a rectangular cake with chocolate toppings as displayed in Fig.~\ref{fig_casestudy_cakecut}.
The toppings have two features: shape (stars and hearts), and flavour (white and dark chocolate).
The cake can only be cut into three pieces.
This cake must be distributed to two individuals A and B, that differ in their contributions to making or paying the cake, and preferences on topping features. 
They have in common, that they value the amount of cake similarly.

The individual's utility function for the received pieces of cake is the sum of three components:
\begin{itemize}
    \item amount of cake (full cake equals one unit of utility)
    \item utility points according to topping shape
    \item utility points according to topping flavour
\end{itemize}

There are eight different ways to distribute the three pieces amongst the two individuals (assuming that no piece is wasted).
The quantitative, dianemetic fairness measures enables the objective analysis of this problem, and to fairness-optimally allocate the resource amongst the individuals.
Following measures are used for the different guiding principles: minimum utility (difference principle), standard deviation of utility (equality principle), standard deviation of contributions (equality-of-opportunity principle), sum of utilities (greater-good principle), standard deviation of utility-contribution ratios (proportion principle), threshold share of utility, with 0.50 utility as a sufficient minimum utility (sufficiency principle).

Let us discuss one of the eight possible allocations to outline the calculations.
In this case study, we leave the contributions constant, meaning A contributes 0.9 X to the cake, and B contributes 0.1 X to the cake. 
Assume A gets piece 1 and 3, and B gets piece 2 (allocation scenario 4), then A receives 0.8 Y, B receives 0.2 Y.
The utility the individuals experience based on the cake they get depends on the amount Y and the toppings. 
A experiences a utility of $U = 1.0 = 0.8 + 2*0.1$, as A receives 0.8 amount of cake, and additionally two toppings in heart and star form which both add 0.1 up to the utility; as A is interested in dark chocolate only but pieces 1 and 3 are of white chocolate, no additional utility can be generated for A. 
B experiences a utility of $U = 0.5 = 0.2 + 0.2 + 0.1$, as B receives 0,2 amount of cake, and additionally one topping in heart form which adds 0.1 up to the utility, and the topping is made of dark chocolate which adds 0.2 up to the utility.

The quantitative fairness assessment of the different resource allocation scenarios reveals different levels of fairness depending on the different guiding principles. 
Following the difference principle, scenario 5 achieves the highest worst case with a utility of 0.7.
Following the equality principle, scenario 3 achieves the lowest inequality / dispersion (standard deviation) in utilities.
A discussion of the equality-of-opportunity in this case study might not be relevant, as we did not provide details on the reasons for contributions.
Following the greater-good principle, scenario 5 achieves the highest total utility.
Following the proportion principle, scenario 1 achieves the highest alignment with contributions.
Following the sufficiency principle, scenarios 3-5 provide a sufficient minimum experienced utility for both individuals.

These differing ratings for fairness by the different guiding principles highlight the conflicting differences in understanding of what fairness is.
Scenario 5 for instance, achieves the best situation in total, and even for the poorest, while it is twice as unequal when compared to scenario 3. 
From a purely merits based perspective, A should receive everything as in scenario 1, as A contributes almost everything for buying the cake.

\subsection{Fishermen \& diorthotic fairness}
Let us discuss diorthotic fairness at the example of the daily work of two fishermen, as shown in Fig.~\ref{fig_casestudy_fishing}. 
Assume there is a village with two fishermen, that go to the same pond for fishing every day.
The fishermen differ in their capabilities due to experience and catching methods they use, but also in their working attitude and willingness to work.
Should each fisherman keep what they catched, or should the village agree on some form of legislation to distribute the daily catch "fairly" to both fishermen?

This example is a metaphor for various diorthotic fairness contexts:
\begin{itemize}
    \item Imagine a nation's labor economy. People have different capabilities to earn an income, different chances depending on the economic situation of markets, exports, industries, and also different levels of discipline and perseverance. A government could apply an income taxation to redistribute wealth amongst the individuals of the labor force. This could create an overall healthier economy, as workers with bad luck or health in one season can still survive based on the solidarity support of others. However, this could also create a demotivation of the top performers, as higher achievements mean higher contributions to the others, that might not work as hard as them.
    \item Imagine a nation's public health system. People have different chances to get sick based on genetic and socio-economic dispositions that they cant affect, but also do they make life choices that lead to different health-affecting life-styles (e.g. smoking). Should the government enforce a mandatory health insurance system, so that everyone pays into the system, even when not using it? The solidarity could lead to less extreme poverty, as often severe diseases, accompanied by job loss and divorce, lead to homelessness. However, many diseases could be prevented by a healthier life style, and a governmentally-enforced safety net could incentivize a free-riding behaviour, in which individuals have even less incentivizes to take care of them, as they can get cured on the costs of others. If such a public health system and insurance would be implemented, how should people pay for it? Based on their income, based on their lifestyle, or even based on their genetic and socio-economic dispositions?
\end{itemize}

For this case study, let us consider two fishermen, A and B.
Fisherman A works 8 hours per day, and has an equal chance to find 0 to 6 fishes per day. This means, on average, A catches 3 fishes per working day, and 0.375 fishes per hour.
Another fisherman B works 12 hours per day, and due to his experience and catching method, catches exactly 5 fishes per day. This means, on average B catches 5 fishes per working day, and 0.417 fishes per hour.

Let us assume one day, A catches 2 fishes, and B catches 5 fishes, so there are 7 fishes in total to be distributed.
As fish is a perishable good, it does not make sense to not allocate all fishes, so the sum of all fishes of A and B (5+2=7) forms the Pareto-efficient frontier.
Moreover, each fisher has the same utility for the amount of fish received, however when walking back home, fisher A looses around 5\%, and fisher B looses around 15\%.

Fish can be considered as a divisible good, one could assume continuous distribution based on weight. This means there is an infinitely large solution space for allocations along the Pareto-efficient frontier.
The quantitative, diorthotic fairness measures enable us to objectively analyse this problem and to fairness-optimally allocate the resources. 
Following measures are used for the different guiding principles: Rawlsian welfare function (difference principle), Foster welfare function (equity principle), standard deviation of working time (equality-of-opportunity principle), sum of all utilities (greater-good principle), standard deviation of received fish-working time ratios (proportion principle), threshold share of received fish, with 2 units as a sufficient minimum (sufficiency principle).

The quantitative fairness assessment of the different resource allocation scenarios reveals different levels of fairness depending on the different guiding principle of fairness.
Following the difference principle, it would be most fair to equally distribute the catched fish, meaning both fishermen get 3.5 units of fish.
Following the equality principle, it would be most fair to equally distribute as well.
A discussion of the equality-of-opportunity in this case study might not be relevant, as we did not provide details on the reasons for working times. It might be, that A is older and cannot work that long any more as B does. In this case, one would need to discuss ways to account for that.
Following the greater-good principle all fish should be given to A, as A can bring most of fish home where it can actually be cooked, while giving fish to B would generate slightly more waste. If the greater-good principle would be applied to resources y rather than utilities, then the greater-good principle would be indifferent on any allocation along the Pareto-efficient frontier.
Following the proportion principle, A should receive 2.8 and B should receive 4.2 units of fish, according to their contributed working time. One could argue, that not working time, but actual contribution matters as Aristotle argues; in this case each fisher should keep as much as they catched without sharing.
The sufficiency principle would be indifferent to all allocations along the Pareto-efficient front between 2 to 5 fishes for A resp. B.

The different social welfare functions exhibit different shapes. 
Difference and equality principle share the same optima and similar gradients.
The greater-good principle is almost parallel, and the proportion principle is even almost orthogonal to the Pareto-efficient front.
The sufficiency principle rather distributes the solution space into distinct areas. 

\subsection{Holistic, ideology-agnostic fairness discussions}
The proposed quantitative framework purposely does not advocate one over another guiding principle, but rather enables a holistic, and integrative analysis.
Deviating from the fairness-optimal allocations from one guiding principle by just a bit, can achieve significant improvements form the perspective of another guiding principle.
A compromise solution might be derived based on the preferences and weights the decision maker provides to the different guiding principles.
Another way could be: rankings of all alternatives based on the different guiding principles could be created, and then aggregated to a combined ranking for a final decision making.

\subsection{Transitive and quantitative, rather than transcendental and normative fairness discussions}
Contrary to previous, rather transcendental, qualitative and mostly normative, philosophic works, our framework can be used to analyse situations from a more transitive, and quantitative perspective.
Not only can two situations be compared to decide which one is more fair (transitive), but also can the framework be used to assess how much more fair it is (quantitative).
This enables a more systematic discussion of the deviation from strict optima, and encourages an inclusive discussion that allows for the combination of different goals, including different fairness and efficiency definitions. 
Similar to quantitative definitions of efficiency, quantitative definitions of fairness can be used as a goal metric for the design and optimization of algorithms.

\subsection{Equality-of-opportunities guiding principle and fair resource allocating algorithms}
The equality-of-opportunities guiding principle is related to the chance aspect of procedural fairness and a guiding principle for the distributive fairness. 
In the case studies we excluded a discussion, as further assumptions must be taken. 
Besides, the question that needs to be answered for a discussion from this guiding principles perspective includes how the resource allocation can affect the opportunities and chances individuals have. 

If there is a clear relationship between the allocated resources in a cycle, and the chances of individuals in the next cycles, then equality-of-opportunity is clearly relevant to discussing the allocation of resources.
For instance, one could assume that having more money at the beginning of market opening will allow to generate more trading profits, which will then enable even more chances and opportunities at the beginning of the next day.
This could be considered as a feedback loop, and therefore redistribution of shares of profits that are due to luck rather than capabilities or efforts, might be considered as fair, when they enable more chances for everyone else.

If there is no clear relationship between the allocated resources and the chances, a different discussion is necessary (e.g. how much fish you get as fisherman in the case study wont affect how much fish you can get on the next day).
One could rather focus on which decision criteria an algorithm uses, or how it weights different inputs to the decision making.
Doing so, one could aim for inputs that reflect more equal opportunities for individuals to participate and actually affect the outcome.

Rather than focusing on pure contribution of making or buying the cake in the dianemetic case study above, one could try to adjust and normalize the inputs for capability (how much pocket money do you have available to pay for the cake) or ability (how much knowledge and tools do you have to make a cake).
Rather than focusing on pure contribution of fish, or working time in the diorthotic case study above, one could try to adjust and normalize these inputs for capability (catching rate of fish) or ability (age, gender, size), to better reflect the pure willingness to work.

\section{Conclusion} \label{conclusion}

This work set out to propose an useful, holistic, quantitative, transactional, distributive fairness framework, which enables the systematic design of socially-feasible decision-making systems in the context of equitable cybernetic societies.

After the review of distributive and procedural fairness theories from relevant literature and domains, algorithmic fairness and the importance of transparency and explainable AI were elaborated.

The proposed quantitative fairness framework offers measures for dianemetic and diorthotic fairness discussions based on statistic metrics, dispersion metrics, and social welfare functions.
Two case studies on fair cake-cutting and fishermen demonstrate the usefulness and flexibility of the proposed framework. 

Future work could focus not only on situational quantification of fairness at a specific time, but to include a temporal component for repeated settings.
For example, forms of aggregation over many iterations of the same algorithm could be part of investigation. A useful way could be the probabilistic, stochastic discussion of the distributive effects of algorithms.

\backmatter






\bmhead{Declaration of competing interest}
None.

\setlength{\bibsep}{0pt plus 0.3ex}
{\tiny \bibliography{sn-bibliography} }

\end{document}